# A Seamless Flow Mobility Management Architecture for Vehicular Communication Networks

Rodolfo I. Meneguette, Luiz F. Bittencourt, and Edmundo R. M. Madeira

***Abstract:*** **Vehicular Ad Hoc Networks (VANETs) are self-organizing, self-healing networks which provide wireless communication among vehicular and roadside devices. Applications in such networks can take advantage of the use of simultaneous connections, thereby maximizing the throughput and lowering latency. In order to take advantage of all radio interfaces of the vehicle and to provide good quality of service for vehicular applications, we developed a seamless flow mobility management architecture based on vehicular network application classes with network-based mobility management. Our goal is to minimize the time of flow connection exchange in order to comply with the minimum requirements of vehicular application classes, as well as to maximize their throughput. NS3 simulations were performed to analyse the behaviour of our architecture by comparing it with other three scenarios. As a result of this work, we observed that the proposed architecture presented a low handover time, with lower packet loss and lower delay.**

## I. INTRODUCTION

Vehicular Ad Hoc Network (VANET) is a rising subclass of Mobile Ad Hoc Networks which provides wireless communication among vehicles as well as between vehicles and roadside devices. In VANETS, each vehicle can have multiple radio interfaces, thus being able to simultaneously connect to different domains and radio access network technologies [1]. Although these vehicles can connect to these different network technologies simultaneously, nowadays vehicles are limited to choosing a default interface for sending and receiving information. This limitation is related to the current model of multiple interface management, where several interfaces are attached to the operating system [2]. Usually, operating systems use configuration files from the user, or consider the types of applications to select a default network interface to send and receive data [3].

To allow the use of more than one network interface simultaneously, the Internet Engineering Task Force (IETF) has developed the technology of IP flow mobility, which can divide IP flows among multiple links according to application requirements and user preferences. There are some groups of the IETF, such as the IETF mobility extensions for IPv6 (MEXT) [4] and IETF-based network mobility extensions (NETEXT) [5], which have been working on the development and elaboration of a pro-tocol that allows the use of more than one interface simultaneously. The MEXT has standardized a host-based IP flow mobility in Mobile IPv6, enabling flow bindings for the mobile node with multiple interfaces [4]. This method has an air waste resource problem, such as establishing IP-in-IP bi-directional tunnel over the air interface and exchanging mobility-related layer 3 (L3) signaling messages via the wireless link [6]. To avoid overloading the wireless network, the NETEXT discusses the use of a network-based IP flow mobility in Proxy Mobile IPv6 (PMIPv6). This solution has the limitation where the flow mobility of the mobile nodes (MNs) should be initiated and controlled only by network-side entities [6]. Moreover, there exist some limitations on flow mobility support [7]. In order to contour these limitations and to enable the dynamic movement of individual flows according to the flow control, we propose a new mobile-initiated seamless IP flow mobility mechanism, developed over the network-based mobility management architecture.

In this paper we propose and evaluate a seamless flow mobility management architecture (SFMMA) in vehicular communication networks. The proposed architecture deals with different network interfaces at the same time, seeking the maximization of the network throughput and keeping low latency and low packet loss rate. To achieve this, our scheduling considers that applications are divided into three classes, according to general goals of vehicular network applications: safety, comfort, and user. Moreover, this model considers that vehicles are moving into a city or in a highway, and that people in the vehicles are running more than one application class at the same time.

This paper is organized as follows. Section II introduces the terminology and basic concepts involved in the paper subject. Section III discusses some related works. Section IV presents the proposed seamless flow mobility management architecture in vehicular communication networks, while in Section V we present and analyze simulation results. Section VI concludes the paper with remarks and future directions.

## II. BACKGROUND

This section presents some basic concepts involved in this paper, introducing VANETs and mobility management.

### A. VANET

Vehicular Ad Hoc Networks (VANETs) are aimed at communication between vehicles and / or between vehicles and roadside infrastructure [8]. They can use cell phone towers or even an outside access bridge for such communication. In 1999, the Federal Communications Commission (FCC) allocated a frequency spectrum for inter-vehicle communication and between

R. I. Meneguette, L. F. Bittencourt, and E. R. M. Madeira are with the Institute of Computing (IC), University of Campinas (UNICAMP), Campinas, São Paulo, Brazil Email:{ripolito, edmundo, bit}@ic.unicamp.br

vehicles and roadside infrastructure, establishing rules and licensing services for the Dedicated Short Range Communications (DSRC) at the 5.9GHz band [9]. This protocol is an extension of IEEE 802.11, the 802.11p, being a technology for the vehicle environment at high speed. The physical layer (PHY) is adapted from the IEEE 802.11a PHY, and the multiple access control (MAC) layer is very similar to the IEEE 802.11 MAC [10].

In VANETs, devices may suffer frequent disconnections and access point changes on their route due to: (i) low network data transmission rate; (ii) the high speed that vehicles can achieve on a highway; and (iii) decision-making that changes the device's route [11]. Therefore, it is currently a challenge to smoothly change access points in a way that the user does not notice any change in the application performance [12].

### B. 802.21 Protocol

The IEEE 802.21 [13] provides a standard to assist the implementation of vertical handovers, and it is a recent effort to allow the transfer and interoperability between heterogeneous network types. The goal of IEEE 802.21 is to improve and facilitate the use of mobile nodes, providing uninterrupted transmission in heterogeneous networks. The most important tasks of the IEEE 802.21 are the discovery of new networks in the environment and the selection of the most appropriate network for a given need. The network discovery and process selection are facilitated by network information exchange that helps the mobile device to determine which networks are active in its neighborhood, thus allowing the mobile device to connect to the most appropriate network based on its own policies [14].

The core of the 802.21 is the Media Independent Handover Function (MIHF). The MIHF has to be implemented in all devices compatible with the IEEE 802.21 (in hardware or software). This function is responsible for communicating with different terminals, networks and remote MIHFs, and also for providing information services to the higher layers [15]. The MIHF defines three different services: Media Independent Event Service (MIES), Media Independent Command Service (MICS), and Media Independent Information Service (MIIS). These services allow obtaining and storing relevant information about the network status such as loss, throughput, and what are the subnets. Part of this information is found on the information elements IE_CONTAINER_NETWORK and IE_CONTAINER_POA of the protocol architecture, which are used in the context of our work. For more details on the 802.21 protocol, please refer to [13].

### C. Proxy Mobile IPv6

Proxy Mobile IPv6 (PMIPv6), as specified in [16], provides network-based mobility management to hosts connecting to a PMIPv6 domain. PMIPv6 introduces two new functional entities, namely the Local Mobility Anchor (LMA) and the Mobility Access Gateway (MAG). The MAG is the first layer three hop detecting Mobile Node attachment and providing IP connectivity. The LMA is the entity assigning one or more Home Network Prefixes (HNPs) to the MN and is the topological anchor for all traffic from/to the MN. The fundamental foundation of PMIPv6 is based on MIPv6 in the sense that it extends

MIPv6 signaling and reuses many concepts such as the home agent (HA) functionality. The LMA and the MAG establish a bidirectional tunnel for forwarding all data traffic belonging to the Mobile Nodes. The network-based localized mobility support provided by PMIPv6 was designed for hosts, so a mobile host can freely roam within the PMIPv6 domain, without changing its IP address [17].

Figure 1 shows a simple topology of the PMIPv6 protocol. We have a mobile node (MN), a MAG connected to a wireless access point, and an LMA. We also have an LMA connected to a corresponding node (CN), which can be any node on the Internet or in the LMA that communicates with the MN.

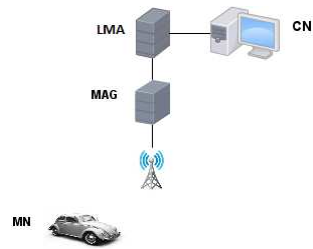

Fig. 1. PMIPv6 topology and components.

## III. RELATED WORK

This section presents some proposals related to mechanisms that, in some sense, enhance handover performance by enabling the simultaneous use of multiple interfaces during flow mobility.

Choi et al. [7] proposed a flow mobility support for updating and full-covering of the flow mobility support in PMIPv6. This proposal is based on a virtual interface in the mobile network. A virtual interface makes all physical interfaces to be hidden from the network layer and above. Flow Interface Manager is placed at the virtual interface, and Flow Binding Manager in the LMA is paired with Flow Interface Manager. They manage the flow bindings and are used to select the proper access technology to send packets. Flow mobility procedure begins with three different triggering cases, which are caused by a new connection from the MN, by the LMA's decision, or by a request from the MN.

Melia et al. [18] focused on the design and implementation of flow mobility extensions for PMIPv6. To do this, the authors extended PMIPv6 to support dynamic IP flow mobility management across access wireless networks according to operator policies. Considering energy consumption as a critical aspect for hand-held devices and smart-phones, they assessed the feasibility of the proposed solution and provided an experimental analysis showing the cost (in terms of energy consumption) of simultaneous packet transmission/reception using multiple network interfaces. In their proposal, the network (in particular the mobility anchor) is the decision control entity.

Kim et al. [6] proposed an optimized seamless IP flow handover mechanism for traffic offloading, mobile-initiated with network-based mobility management. The proposed mechanism defines new mobility headers, including a handover initiation for flow mobility (HIF) and a handover acknowledge for flow mobility (HAF) messages, which are extended from a handover initiate (HI) and a handover acknowledge (HAck), for efficient support of mobile-initiated flow mobility in PMIPv6. In addition, a new mobility option is defined for conveying interface action information, named interface-status-and-action (ISA) option. In order to minimize the end-to-end packet delay and maximize the throughput during the IP flow handover, no bi-directional tunnel is established between the MAGs.

Table 1. Related work characteristics.

| Work | Protocol | Auxiliary | Decision | Technology | State |
|------|----------|-----------|----------|------------|-------|
| Choi [7] | PMIPv6 | HUR, HAR, Flow mobility messages | LMA, MN | Wlan, WiMAX, PPP(3G) | No |
| Melia [18] | PMIPv6 | - | LMA | Wifi, 3G | No |
| Kim [6] | PMIPv6 | HIF, HAF, ISA messages | MN | Wifi, 3G | No |
| Makaya [3] | MSM | MIHF | MN | Wifi, 3G | Yes |
| Proposal (SFMMA) | PMIPv6 | MIHF | MN, MAG, LMA | LTE, 802.11p | Yes |

Makaya et al. [3] proposed a new mechanism for selective IP traffic offload (SIPTO) for vehicular communication networks. This mechanism provides support for data offload, seamless handover, and IP flow mobility for mobile devices equipped with multiple interfaces. The authors created a multilink striping management mechanism that allows data offload and flow mobility across different access network technologies. The reports on link quality and network (i.e., core and access) status are used as the triggers for the MSM to decide whether flow mobility, data offload or handover is needed to avoid session disruptions. The MIH services are used to trigger the need of IP flow mobility, data offload, or handover. By using the MIH primitives, the IP flow mobility, handover, and data offload are done seamlessly and allow a better usage of network resources while enhancing network capacity.

Table 1 summarizes the following characteristics of the related work and the proposed seamless flow mobility management architecture: (i) the technology used for mobility management (protocol); (ii) the use of other protocol or messaging auxiliary (auxiliary); (iii) the device used to initialize or to make decisions on flow changes (decision); (iv) which network technology it has been tested into (technology); and (v) if it takes into account the state of the network and its flows (state).

Our architecture is based on application classes of vehicular networks with network-based mobility management. Not only the Local Mobility Anchor (LMA) and the Mobile Node (MN) can initiate the change of flow, as a Mobile Access Gateway (MAG) also has the ability to start the change of flow, making these changes more dynamic. Furthermore our architecture takes into account not only the state of the network, but also the state of each flow, providing to the node relevant information about the flow to make the best decision at the time of the flow exchange.

## IV. A SEAMLESS FLOW MOBILITY MANAGEMENT ARCHITECTURE

The proposed Seamless Flow Mobility Management Architecture – SFMMA – considers a common infrastructure for multiple access technologies in a transparent way. We created a multi-access wireless network architecture using the Wi-Fi, Vehicular, WiMAX, and LTE technologies, providing a continuous and transparent connection for the vehicular applications.

The objectives of the architecture are to maximize network throughput and to keep latency and packet loss within the minimum requirements for vehicular applications. To accomplish this, the proposal creates a flow mobility management based on application classes of vehicular network and on the status of each active network on the environment.

Our architecture differs from other architectures for vehicular networks because it considers the needs of each application class, such as throughput, delay and packet loss. Moreover, it uses the state of the currently active network to perform the division of the flow. Another difference is that not only the LMA can take the decision to split the flow, but also MAG and MN can take action in the flow divisions.

### A. OVERVIEW OF THE PROPOSED ARCHITECTURE

The SFMMA is divided into three modules: one module that acts in MAG, one in LMA (Figure 2(a)), and another module that is embedded into the vehicle (Figure 2(b)).

In the MAG and LMA architecture (Figure 2(a)), the network layer contains the PMIPv6 protocol for handling node addresses and the prefix of the access point required for routing messages. The MAG and the LMA architecture also has an *MIHF* module that has the extended functions of the 802.21 protocol. The difference between MAG and LMA architectures is the information that is stored by the *IS Network service*, where LMA has an overview of the state of the network connected to it, whilst MAG has a local view, since it only monitors its own links.

The vehicle architecture contains a mobility management module, called Vehicle MIH, which comprises a requirement management module that receives the minimum network requirements for the application to run. The Vehicle MIH module (Figure 2(b)) also has a module for link selection, which receives the network status information and decides whether it will perform a handover and, if so, to which network it should connect. Besides this, the link selection module helps the logical interface to decide which link will be used to send a particular message. Both the requirement management module and the link selector module send commands to the MIHF module. The MIHF module is an extension of the functions of the 802.21 protocol.

In order to balance the flow of information among networks, it is necessary to extend the 802.21 protocol. This extension enables the creation of a flow manager based on characteristics of each application of vehicular networks, such as delay and throughput, and makes the node to have more knowledge about which network interfaces are active at a given time.

### B. FLOW DEFINITION

In our work, we grouped applications into three classes according to general goals: safety, comfort, and user. The *safety* class comprises applications aimed at assisting drivers in handling unpredictable events or dangers by actively monitoring nearby traffic environment through message exchanges [19]. The *comfort* class comprises applications that focus on comfort and efficiency of the street or road. In other words, these applications enhance the degree of convenience of drivers and traffic efficiency by sharing traffic information among roadway infrastructure and ongoing vehicles [19]. The *user* class is composed of applications which are concerned with providing increased satisfaction to the occupants of the vehicle with information, advertisements, entertainment, and various types of communication services [20].

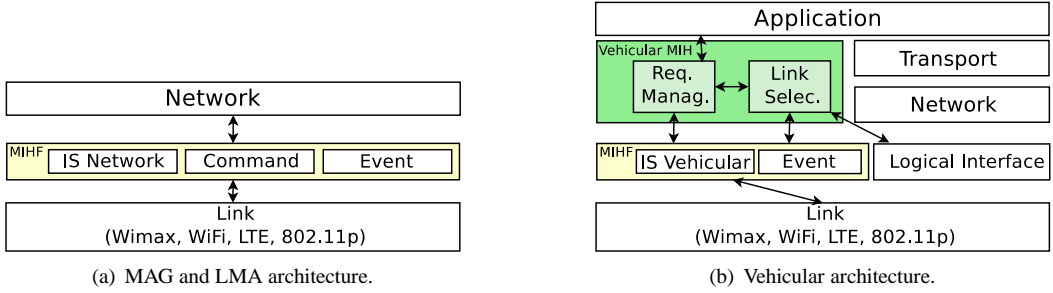

(a) MAG and LMA architecture.

(b) Vehicular architecture.

Fig. 2. Components of the SFMMA.

This classification for vehicular network applications allows a division of information flow in the network in three distinct flows: the first flow is for safety class, the second flow is for comfort class, and the third flow is for user class. To differentiate these flows we used a 2-tuple, which consists of the protocol used to transmit it and the destination port. Due to the variety of applications possible in each application class, we could divide the destination ports into ranges, one for safety class, a second one for comfort class, and a third one for user class. Beyond the flow description, other parameters are required to perform the division of flow such as a flow identifier, the priority of each flow, and to which interface it is connected. An abstraction of these parameters is presented in the Table 2.

Table 2. Abstraction of flow parameters.

| ID-Flow | Priority | Flow Description | ID-Interface |
|---------|----------|------------------|--------------|
| 1 | 1 | (UDP or TCP , 5001 - 5100) | 1, 2 |
| 2 | 2 | (UDP or TCP , 5101 - 5200) | 2 |
| 3 | 2 | (UDP or TCP, 5201 - 5300) | 2 |

The abstraction shown in Table 2 represents what is stored in the information service of vehicles, where the *id_flow* is the identifier of the flow, *priority* is the priority that a flow has over others, the *flow description*, and *ID-Interface* is the interface identification. For the LMA and MAG, instead of having an interface ID, they have an identifier of the binds between the MAG and the LMA.

In order to utilize this flow control structure, it was necessary to extend the 802.21 protocol with knowledge about the new information service.

## C. EXTENDED 802.21 PROTOCOL

The 802.21 protocol services (MIES, MICS and MIIS) allow obtaining and storing relevant information about the network status such as loss, throughput, and what are the subnets. We extend the MIIS so that the node can get flow information and also allow the node to manage the flow division based on the state of the network and the active interfaces.

This extension of the protocol was used in all nodes which participate in the network as mobile nodes, MAG nodes, and LMA nodes. Each type of node has a different extension of the 802.21 protocol, i.e. an extension for the mobile node, another for MAG nodes, and a third one for LMA nodes.

### C.1 802.21 IN THE MOBILE NODE

Besides the MIHF standard features, the mobile node has an extension of the 802.21 protocol. This extension will provide information from the states of network interfaces, i.e. which is the MAG where the network interface is connected, and what are the home network prefixes (HNP). This information is necessary for the node to know whether the interface is active or not, to whom it needs to route the messages, and what are the prefixes that it can use to send its messages. To comprise this extension, we have created three different elements in information elements (IE) of the original protocol, as seen in Figure 3: (i) IE-interface, which manages the interface information; (ii) IE-Flow, which stores information of flow control applications; and (iii) IE-Container-FlowStatus, which stores the state of each flow.

Figure 3(a) describes the IE-interface, containing an interface ID that is the *IE_OPERATOR_ID* of the 802.21 standard protocol. It identifies the network interface, which is the technology being used, the cost of using this technology, and other information. This identification and interface information are contained in *IE_CONTAINER_NETWORK*. The MAG field contains the identifier of the POA information that is linked to the interface at that moment. This information is contained in *IE_CONTAINER_POA*, which contains information from every possible POA that the interface can be connected. HNP contains a list of prefixes to which the MAG (POA) can forward data, and the *interface status* representing if the interface is active or inactive. Figure 3(b) describes the IE-Flow containing information of the flow control described above. The IE-FlowStatus (Figure 3(c)) contains parameters such as throughput, delay, and packet loss, with their respective values for each flow.

This information is replicated to the link selector, thereby giving it greater flexibility in the decision to change the flow from one interface to another, as well as to expedite the choice of interface to send a particular message.

### C.2 802.21 IN THE MAG NODE

As the mobile node, the node MAG also has the 802.21 protocol standard features and the IE-flow and IE-FlowStatus. How-

| IE - INTERFACE | | | |
|---|---|---|---|
| ID - Interface | HNP | MAG | Status |
| 1 | HNP1; HNP2; HNP3 | POA1 | ACTIVE |
| 2 | HNP2; HNP3 | POA2 | ACTIVE |

(a) IE-interface

| IE - FLOW | | | |
|---|---|---|---|
| ID - FLOWSTATUS | Priority | Flow Description | ID- Interface |
| 1 | 1 | (UDP or TCP, 5001-5100) | 1, 2 |
| 2 | 2 | (UDP or TCP, 5101-5200) | 2, 1 |
| 3 | 2 | (UDP or TCP, 5201-5300) | 2, 1 |

(b) IE-Flow

| IE - FLOWSTATUS - ID 1 | |
|---|---|
| IE - FLOWSTATUS - ID 2 | |
| IE - FLOWSTATUS - ID 3 | |
| PARAMETERS | VALUE |
| THROUGHPUT | |
| PACKET LOSS | |
| DELAY | |
| ⋮ | |

(c) IE-Container-FlowStatus

Fig. 3. Information elements introduced in the proposed architecture.

ever, the IE-flow of the MAG does not have the interface ID. Instead, it has the identifier of the Binding Cache Entry between MAG and the LMA node, as seen in Figure 4.

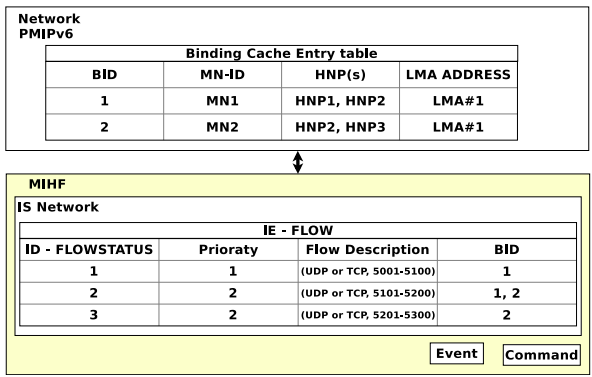

Fig. 4. MAG Architecture.

Figure 4 shows an abstraction of the information stored in the PMIPv6 implemented in the MAG node, the Binding Cache Entry table, as well as flow control information stored in the MIHF.

The use of IE-FlowStatus gives the MAG node the power to initiate a possible flow exchange, because the node knows the status of each flow. Therefore, if some parameter of the IE-FlowStatus is greater than an upper threshold, the node can start the change to another MAG. On the other hand, if these values are below a lower threshold, the node could ask for changes in the flow so messages could start to flow through that node. With this, if some parameter of the IE-FlowStatus is greater than an acceptable upper or lower limit, the node can request changes to be applied to the flow.

### C.3  802.21 IN THE LMA NODE

The LMA node also has the standard functionality of the MIHF, but as in MAG, the LMA node has the information flow control and abstraction of flow state. This abstraction refers to the flow state of each MAG, i.e., it was necessary to isolate the information flows of MAGs that are within the LMA domain to give greater flexibility in flow decision-making. To do this, we created a new element, *IE-CONTAINER-MAG*, which is an extension of *IE_CONTAINER_POA*, i.e., the *IE-CONTAINER-MAG* contains all the information about the POA *IE_CONTAINER_POA*. Moreover, it provides information for the flow state of each POA, as we see in the Figure 5. The *IE-CONTAINER-MAG* contains only information of MAGs (POA) that are managed by LMA.

| IE_POA_LINK_ADDR |
|---|
| IE_POA_LOCATION |
| IE_POA_CHANNEL_RANGE |
| IE_POA_SYSTEM_INFO |
| IE_POA_SUBNET_INFO |
| IE-Container-FlowStatus |
| IE_POA_IP_ADDR |

Fig. 5. IE-CONTAINER-MAG definition.

With this, the LMA node has an overview of the flow states across the network, and it is able to decide to change a flow from one technology to another, or even to change the path of a flow at any moment.

### D.  HANDOVER PROCEDURE

Generally, the formation of a flow occurs when a mobile node starts an application and starts sending messages to the network. However, the way that the flow goes through the network can be changed over time, i.e., along its existence, a flow which was originally set to be sent by a particular network technology can be routed to another technology. This change in the flow can be initialized in three different locations: in the mobile node, in the LMA node, or in the MAG node.

### D.1  MOBILE NODE

A flow exchange can be triggered by the mobile node (MN) for two reasons: (i) the activation of a network interface occurs; or (ii) network parameters or the actual flow state is outside the minimum requirements of the application (throughput, packet loss, and delay).

When a new network interface is activated, there are two events to consider: (1) if the mobile node has no flow to that interface, i.e., all applications are mapped to another network interface, and (2) if there is a flow previously mapped to that interface, if this flow has been established in the LMA, it assigns the same home network prefix with the target flow. Otherwise, the LMA assigns new home network prefix.

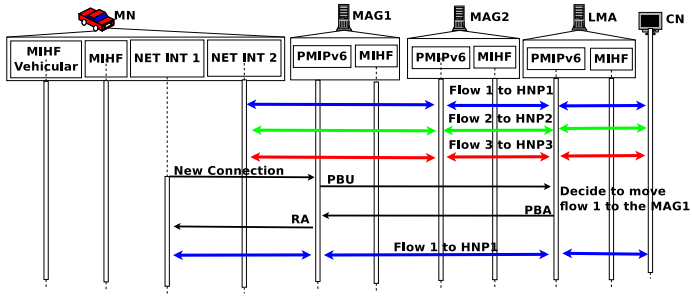

Fig. 6.  Sequence Diagram for flow exchange due to activation of a network interface.

Figure 6 shows the flow exchange for the flow mobility by the activation of another network interface. There are three traffic flows going over the interface 1. It is assumed that interface 2 is the higher preferred interface for the flow 1. When the 4G interface is activated, flow 1 will be moved into this new interface. In the binding update process, because a flow handover is expected for the new interface, the LMA assigns HNP1 to the new interface. Therefore, the 4G interface gets HNP1 as a home network prefix.

If one parameter of the network is outside of the expected thresholds (packet loss is high, or the delay of a given flow is high), the node can initiate an exchange of interface for that flow.

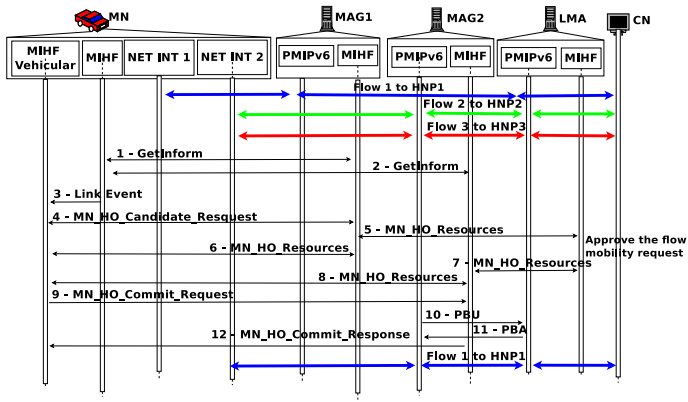

Fig. 7.  Sequence Diagram for flow exchange due to mobile node initiative.

Figure 7 describes the steps involved in the flow exchange when the initiative comes from the mobile node. When the node notices that some of the parameters of the flow state or the network parameters are outside the expected values (steps 1-3) it initiates the flow exchange on that interface. First, the MN sends a request to the MAG where the node is connected (4). This MAG in its turn forwards this request to LMA, which verifies among the MAGs managed by it which one fits the desired features, and then responds to the MAG (5-8). In case there are more than one option that fit into the MAG requests, the LMA may divide these replies by sending some of them to the MAG which is sending the request, and other replies to another MAG to which the node is connected. After this step, the node decides to which MAG the node will connect and sends a request to this MAG (9). The MAG with the LMA performs the update process in its Binding Cache Entry (10-11). Finally, the MAG responds to requests for amendments to the MN. The MN, after receiving this message, starts sending messages for that flow through the new path.

### D.2  MAG NODE

The initiative to perform a flow exchange can also come from MAG, because this node also has the ability to observe the condition of the flow that passes through it. Therefore, the MAG can verify the related thresholds to determine if the network conditions are capable of attending the applications. If some flow exceeds a given threshold, the MAG can communicate the LMA for a possible exchange of flow. This change can occur for two reasons: (i) the status of a flow is below an acceptable lower limit, indicating that the node can handle more traffic, or (ii) the state in a flow is above an acceptable upper limit, indicating an overload in the MAG.

In both cases the procedure for the exchange of flow is the same but with different directions of the flow. In the first case, as the MAG may be experiencing some idleness, it indicates to the LMA that it can route more packets of this flow. In the second case, when the MAG is overloaded, it indicates to the LMA to change that flow to another MAG, thereby relieving the MAG and consequently seeking a better quality in the transmission of that flow. Although there are two separate cases, the way the exchange of flow occurs is the same.

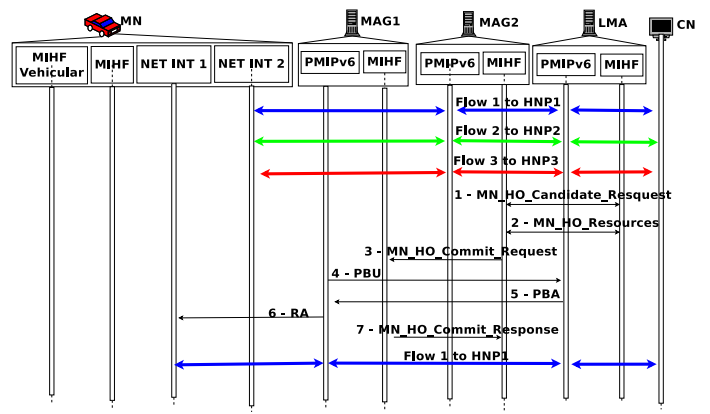

Fig. 8.  Sequence diagram of flow exchange by MAG initiative.

Figure 8 describes the initiative of MAG in the exchange of flow due to a violation of an upper threshold. First, the MAG sends a request to LMA to search for a possible candidate to direct this flow (1-2). After that, MAG sends a message to the chosen node, informing that it will route the flow (3). Then the new MAG, along with the LMA, conducts the whole process of updating the Binding Cache Entry (4-5), and finally updates the information in the MN (6).

### D.3  LMA NODE

The process of flow exchange in the LMA node is much simpler than the MAG and MN because the LMA already has an overview of the current network status, as well as all the information necessary to perform the change. Even so, there are two ways to accomplish this change: (i) when the LMA node knows that MAG has the HNP bound to that flow, and (ii) when the MAG does not have the HNP that is linked to that flow.

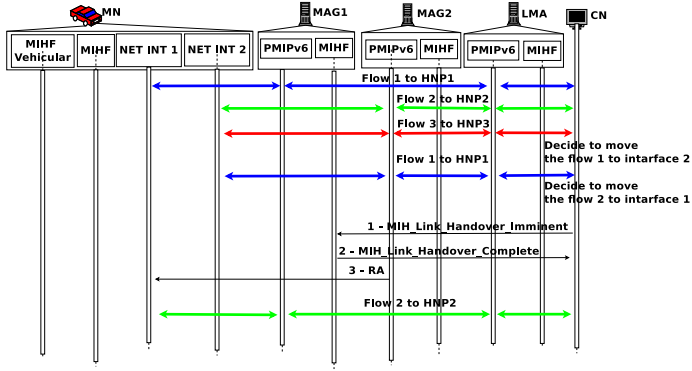

Fig. 9. Sequence diagram of flow exchange by LMA initiative.

Figure 9 describes the change of flow made by the LMA. In the first case, the LMA knows that the MAG2 already has the HNP1 linked to the flow 1, and then the LMA straightforwardly changes the flow. In the second case, MAG1 does not have the HNP2 linked to the flow 2, then the LMA informs the MAG1 that HNP2 will be responsible to forward the packets of flow 2 (1-2). At the end, the MN is notified and begins to transmit packets to the new MAG (3).

## V. SIMULATION AND RESULT ANALYSIS

The proposed seamless flow mobility management architecture (SFMMA) for vehicular communication networks has been implemented in the Network Simulator (NS-3.13). We used the PMIPv6 model that was developed by Hyon-Young Choi [21], as well as the 802.21 model [22]. The purpose of the simulations was to verify the impact that our architecture would cause to both the network and the applications. With that, we aim to verify if the mapping of application classes to different network technologies does not overload the network, if the time of flow change does not affect the application and network, and also to verify if the mapping reaches minimum application requirements such as latency and packet loss. To accomplish that, we used five metrics to evaluate our flow mobility management architecture: throughput, packet loss, delay, delay per application class, and handover time.

In our simulation scenario, each vehicle was running one application of each application class, i.e., one application of safety class, one application of comfort class, and one application of user class. The frequency of messages for each application follows the patterns of the European Telecommunication Standardization Institute (ETSI) [23], where the safety class application sends a message every 0.1s, the user class application sends a message every 1s, and the comfort class application sends a message every 0.5s.

We conducted the simulations with 50 vehicles that were traveling in the map. We then selected a number of vehicles to send and receive messages from 3 different types of application. We varied among 10, 20, 30, 40, and 50 vehicles running the three classes of application at the same time, while the other vehicles only travel creating traffic on the streets but without sending or receiving messages. All vehicles were within the range of a cellular network access point, but the wireless network access points did not cover all the map. The network topology consisted of a wired node, five backbone nodes, one LTE access point, and three 802.11p access points, as illustrated in Figure 10. We performed 10 simulations for each scenario and we computed 95% confidence intervals.

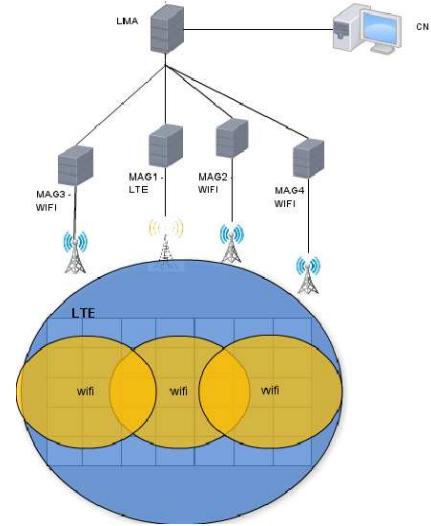

Fig. 10. Simulation Topology.

### A. MOBILITY MODEL

We used two maps to model the mobility of the vehicles. The first one is the Manhattan mobility model and the second one is an excerpt of a real map. To create the Manhattan mobility model we used the bonmotion [24] tools, which created a grid with four rows and six columns. In this map we varied the speed of vehicles among 5, 10, 15, 20, and 25 m/s in order to check if there is any performance impacts on the proposed architecture with different speeds.

As the real map we used a neighborhood in Campinas city, in the state of São Paulo, Brazil. We used the Simulation of Urban Mobility (SUMO) [25] to convert the map extracted from OpenStreetMap [26] to a format that the SUMO simulator could read, as shown in Figure 11(b).

After transforming the real street map to a SUMO map, we used the SUMO tools called *RandomTrips* and *Duarouter* to create random paths for vehicles within the converted map. The speed of the nodes varied from 10 to 16 m/s. Finally, we converted the output of SUMO to NS-3 traces for simulation and result analysis.

### B. RESULT ANALYSIS

We defined 4 different scenarios to evaluate our flow mobility management architecture:

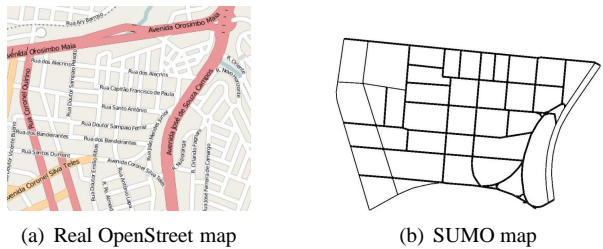

(a) Real OpenStreet map     (b) SUMO map

Fig. 11. Campinas neighborhood map.

- Scenario 0: All nodes use only the LTE network to transmit and receive information, thus no handover is performed.
- Scenario 1: All nodes use only the Wi-Fi network to transmit and receive information, but the nodes can change between Wi-Fi access points, thus performing only horizontal handover.
- Scenario 2 – using the Seamless Flow Mobility Management Architecture (SFMMA): Both LTE and Wi-Fi networks are active in the environment, and nodes can use both networks to send and receive data. In this scenario, the safety application class had preference using the LTE network, and the user and comfort classes had preference in the wireless interface (802.11p). Thus, the safety messages can go through the LTE interface when this interface is up, while user and comfort messages are sent through the 802.11p interface when this interface is up.
- Scenario 3: Both LTE and Wi-Fi networks are active in the environment, but nodes only send and receive information through a single interface, i.e., the interface to which the node is currently connected. All nodes are connected to the Wi-Fi access point at time 0.

For all these scenarios, the data flow is from the vehicle to the wired node. All vehicles have two network interfaces, LTE and 802.11.p. The configuration used for both LTE and 802.11p were the standard for each module in NS-3.13, as defined by procedures such as *wifi.SetStandard(WiFi_PHY_STANDARD_80211p_SCH)*; *WifiMacHelper::Default()*; and *wifiChannel.SetPropagationDelay("ns3::ConstantSpeedPropagationDelayModel")*.

The Manhattan map was used to verify the performance of the proposed architecture by varying the number of participants, and also by varying the vehicles speed. With the Campinas map we aimed at evaluating potential variations in performance in the streets configuration.

Figure 12 shows the average handover time. Scenario 0 was dropped from this analysis for both maps because it does not perform handover. In these graphs we can observe that the SFMMA has a lower handover time in both the real map and Manhattan map. This result is related to the number of handovers occurred and the network status at the time of the handover. Analysing Figure 12(a) for speeds from 20 to 25 m/s, we can observe that in all scenarios there is an increase in the time of handover due to an increase in the number of handovers. However the SFMMA had a better performance because it reduced the number of handovers by taking advantage of previous knowledge about the network conditions and its flows, thereby avoiding unnecessary changes. When analyzing the number of handovers with the real map and with 30 vehicles involved, we observed that the scenario 1 had 18 handovers more than the SFMMA, and the scenario 3 had an average of 8 handovers more than scenario 1 (not shown in the graphs). Thus, scenario 3 had an average of 26 handovers more than the SFMMA. But when comparing the two maps, we observe that the real map had an average handover time 8% faster than in the scenario with Manhattan map.

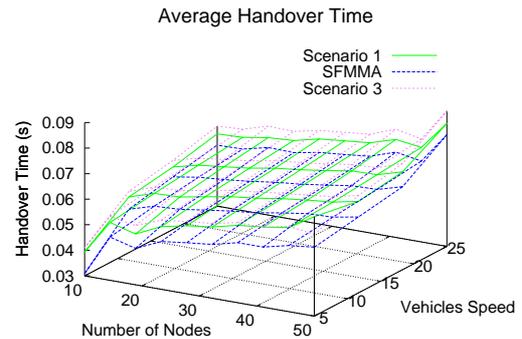

(a) Manhattan Mobility.

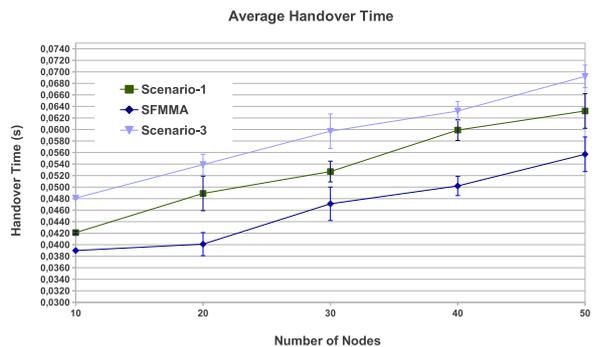

(b) Real Mobility.

Fig. 12. Average Handover Time

We can see from Figure 13 that the SFMMA had fewer packet losses in both maps, showing that the traffic splitted between the two interfaces was effective in avoiding overload in both network technologies. Analyzing Figure 13(a) at speeds above 20-25 m/s all scenarios presented an average of 55% more packet losses compared to lower speeds. This packet loss occurs due to signal interference, added to the time of disconnection of the node, i.e. the handover time, which had an average of 0.065s. Considering 50 vehicles and speed of 10m/s, in the Manhattan map the SFMMA offered an average reduction in packet loss of about 88% when compared to the other scenarios, whilst in the real map the SFMMA obtained an average of 89% of reduction of losses when compared to other scenarios. Moreover, the low rate of packet losses of the SFMMA did not have a relevant impact on the throughput of the network, as shown in Figure 14 and Table 3.

Figure 15 shows the average delay of all application classes. In Figure 15(a) we observe a peak in scenario 1 and in scenario 3. This peak is related to packet loss due to interference and during exchange of access points. With 40 participants and speed of 25m/s, the SFMMA presented a reduction in the average delay of 19% when compared to scenario 1 and 16% when com-pared to scenario 3. In Figure 15(b), when there are 50 cars, the SFMMA had a reduction of 5.5% in average delay when compared to the scenario 3. Besides that, the proposed architecture had an average reduction of 71% over other scenarios in the graph 15(b). The SFMMA performed a better balance for the burden of packets to be sent, as can be inferred from Figure 16.

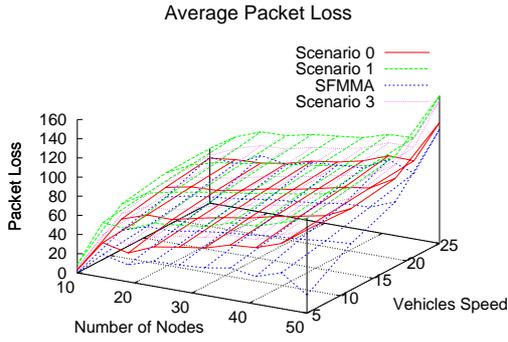

(a) Manhattan Mobility.

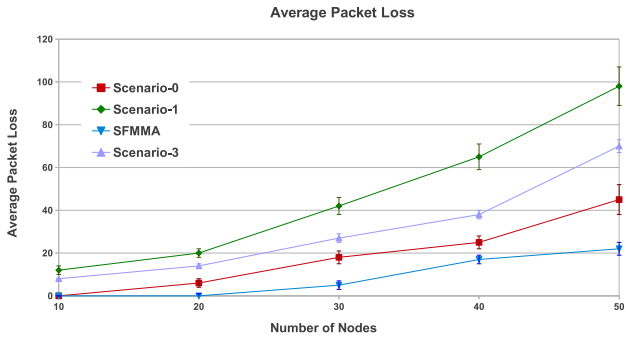

(b) Real Mobility.

Fig. 13. Average Packet Loss

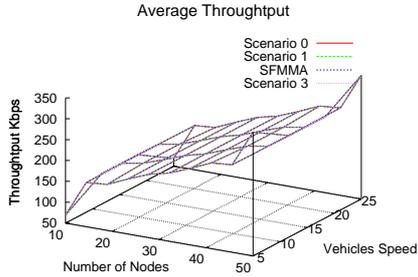

Fig. 14. Throughput for Manhattan map, in Kbps.

Table 3. Throughput for the real map, in Kbps.

| Scenario | Number of Nodes | | | | |
|---|---|---|---|---|---|
| | 10 | 20 | 30 | 40 | 50 |
| Scenario-0 | 69.42 | 138.78 | 208.11 | 277.49 | 346.85 |
| Scenario-1 | 69.31 | 138.65 | 207.91 | 277.20 | 346.34 |
| SFMMA | 69.42 | 138.84 | 208.22 | 277.54 | 346.92 |
| Scenario-3 | 69.35 | 138.71 | 208.05 | 277.39 | 346.54 |

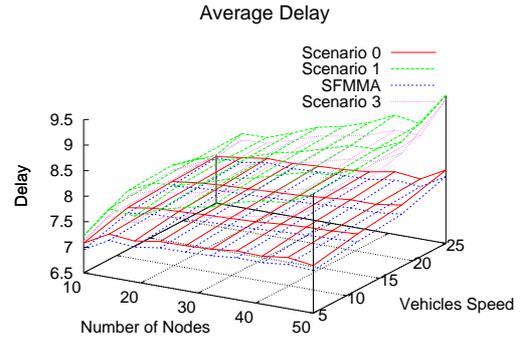

(a) Manhattan Mobility.

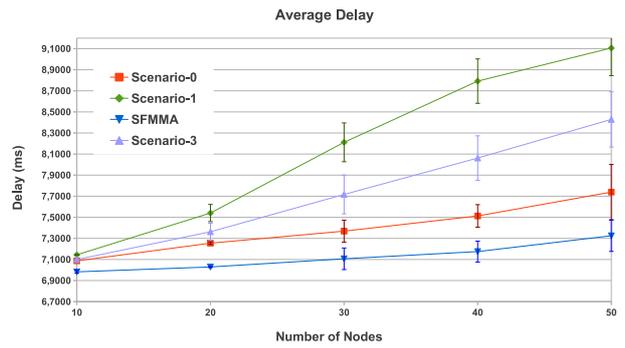

(b) Real Mobility.

Fig. 15. Average Delay

In our proposed architecture, the simulations showed the delay of all applications was granted below the standard values of ETSI [23]. In Figure 16(a), we kept vehicle speed of 10 m/s to obtain the delay per application. We observe from Figures 16(a) and 16(b) that SFMMA achieved a low delay for all classes of application.

Overall, the simulation results suggest that vehicular networks deployed over our proposed flow mobility management architecture present good performance. The proposed architecture achieved an improvement of up to 62% on the average handover time, 88% in packet loss, and 71% in delay, without affecting the throughput. During the simulation, the proposed scenario used all features of the presented architecture, and it achieved low packet loss while maintaining a good network throughput and also a low delay, never exceeding the standards established by ETSI [23]. These results were possible because the proposed architecture is able to divide the burden of packet transmissions among different network interfaces.

## VI. CONCLUSION

In this work we explored the use of more than one network technology to maximize the QoS for applications in vehicular networks. The proposed flow mobility management architecture (SFMMA) deals with different network interfaces at the same time, seeking to maximize network throughput, to decrease the handover time, and to satisfy minimum requirements of latency and packet loss for each class of vehicular network application.

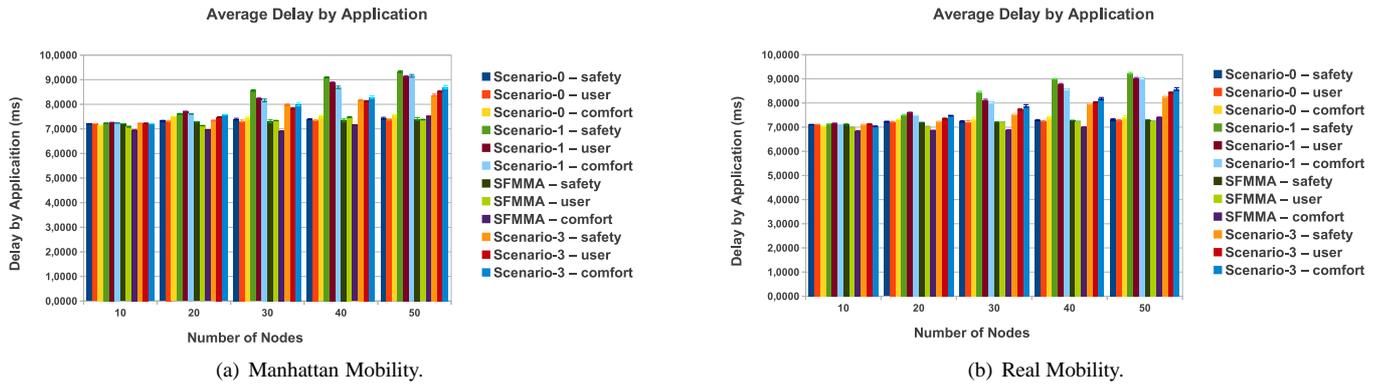

(a) Manhattan Mobility.

(b) Real Mobility.

Fig. 16. Average Delay by Application.

Scenarios where the proposed architecture was utilized presented better performance than common scenarios found in practice. The use of more than one network technology at the same time provided better load balancing in messages to be sent, thereby achieving lower packet losses and shorter delays when there is a large number of participants. Furthermore, no delay in applications exceeded the standard time established by ETSI. The flow control based on the application classes avoided the overload of packets over a single network, thus reducing both the delivery time of messages and the number of lost packets. The results were achieved over two maps of distinct topologies, one representing a city region with standardized squared blocks (known as Manhattan map) and the other one representing a city region with curved streets with no standard blocks.

As future directions, complex mechanisms for network selection are being developed, such as fuzzy networks or Bayes rules, to cope with a larger set of network interfaces.

## ACKNOWLEDGMENTS

We would like to thank CNPq, CAPES, and FAPESP (2009/15008-1) for the financial support.